\documentclass[%
 reprint,
 amsmath,amssymb,
 aps,
]{revtex4-1}

\usepackage{graphicx}
\usepackage{dcolumn}
\usepackage[caption=false]{subfig}
\usepackage[colorlinks,
            linkcolor=blue,
            anchorcolor=blue,
            citecolor=blue
            ]{hyperref}
\usepackage[caption=false]{subfig}
\usepackage{bm}

\renewcommand{\vec}[1]{\mbox{\boldmath$\mathrm{#1}$}}
\def\lb{\left(}
\def\rb{\right)}
\def\ra{\rangle}
\def\la{\langle}
\def\dg{\dagger}

\begin{document}

\preprint{APS/123-QED}

\title{Topological thermal Hall effect driven by fluctuation of spin chirality in frustrated antiferromagnets} 

\author{Yalei Lu$^1$}

\author{Xing Guo$^1$}

\author{Vladimir Koval$^2$}

\author{Chenglong Jia$^{1,3}$}%
\email{cljia@lzu.edu.cn}

\affiliation{$^1$Key Laboratory for Magnetism and Magnetic Materials of MOE, Lanzhou University, 730000 Lanzhou, China \\
$^{2}$ Institute of Materials Research, Slovak Academy of Sciences, Watsonova 47, 04001 Kosice, Slovakia \\
$^{3}$Institut f\"ur Physik, Martin-Luther Universit\"at Halle-Wittenberg, 06099 Halle (Saale), Germany}

\date{\today}
\begin{abstract}
By revealing an underlying relation between the Dzyaloshinskii-Moriya interaction (DMI) and the scalar spin chirality, we have developed the theory of magnon thermal Hall effects 
in antiferromagnetic systems.  The dynamic fluctuation of the scalar chirality is shown to directly respond to the nontrivial topology of magnon bands.  
In materials such as the jarosites compounds KFe$_3$(OH)$_6$(SO$_4$)$_2$ and veseignite BaCu$_3$V$_2$O$_8$(OH)$_2$ in the presence of in-plane DMI,  the time-reversal symmetry can be broken by the fluctuations of scalar chirality even in the case of coplanar $\vec{q}=0$ magnetic configuration.  The spin-wave Hamiltonian is influenced by a fictitious magnetic flux determined by the in-plane DMI.  Topological magnon bands and corresponding nonzero Chern numbers are presented without the need of a canted non-coplanar magnetic ordering. The canting angle dependence of thermal Hall conductivity is discussed in detail as well. These results provide a clear principle of chirality-driven topological effects in antiferromagnetically coupled systems.  
\end{abstract}

\maketitle


\section{Introduction}
Topological excitations of quantum matter are the subject of extensive interest in condensed matter physics. They have been theoretically predicted and experimentally observed in electron systems \cite{PhysRevLett.61.2015,PhysRevLett.95.146802,RevModPhys.82.3045,Roushan:2009kk,RevModPhys.83.1057}. Topological band structure of electrons can be probed by the Hall effect, linked to the Berry curvature throughout the Brillouin zone \cite{RevModPhys.82.1539,RevModPhys.82.1959}. In principle, the concept of topological band theory is independent of the statistical nature of (quasi-)particles. Therefore, the concepts of topological excitations can be extended to neutral bosonic systems such as photons \cite{Lu:2014iu,Khanikaev:2017bq,ozawa2018topological}, phonons \cite{PhysRevLett.95.155901,Kane:2013if,Chen:2014ec,susstrunk2015observation,PhysRevLett.117.068001,Susstrunk:2016df}, and magnons \cite{PhysRevLett.104.066403,Onose297,PhysRevLett.106.197202,Ideue:2012ie,PhysRevB.87.144101,PhysRevB.90.024412,PhysRevB.89.054420,PhysRevLett.115.106603,PhysRevLett.115.147201,2015PhRvB..91l5413L,PhysRevB.91.220408,0953-8984-28-38-386001,Owerre:2016gp,PhysRevLett.117.227201,Li:2016es,PhysRevLett.117.157204,PhysRevB.95.224403,PhysRevB.96.224414,doi:10.1139/cjp-2018-0059}. For spin systems, the magnons do not experience a Lorentz force, which usually drives the electronic Hall effects, but a thermal version of the Hall effect induced by a temperature gradient \cite{PhysRevLett.104.066403}.  
The thermal Hall effect has been confirmed experimentally in insulating pyrochlore \cite{Onose297} and kagome ferromagnets \cite{PhysRevLett.115.106603,PhysRevLett.115.147201}. 

Ordered magnetic insulators with topological magnons as the analogue of Chern insulators requires inherently to break time-reversal symmetry (TRS) \cite{PhysRevLett.61.2015}. In the most cases without applied external magnetic fields, the TRS-broken state is closely related to the (scalar) spin chirality $\hat{\chi}_{ijk}=\mathbf{S}_i\cdot \left(\mathbf{S}_j \times \mathbf{S}_k \right)$ \cite{PhysRevLett.59.2095,PhysRevB.39.11413}. For ferromagnets, Katsura {\it{et al.}} showed that the scalar spin chirality, emerging in the form of ring exchange, provides a fictitious magnetic field for the magnons \cite{PhysRevLett.104.066403}. In the low temperature limit, the thermal Hall conductivity $\kappa_{xy}$ is found to be linearly dependent on the fictitious magnetic flux. In Refs. \cite{2015PhRvB..91l5413L,doi:10.7566/JPSJ.86.011007}, Lee \emph{et al.} generalized the intimate connection between the spin chirality and the Hall-like transport in purely spin systems (including paramagnetic and spin-liquid regimes)  based on the linear response theory. 
On the other hand, for antiferromagnets such as frustrated kagome magnets, a \emph{noncoplanar} spin configuration with \emph{finite} averaged scalar spin chirality $\la\hat\chi_{ijk}\ra$  
{breaks} TRS spontaneously and macroscopically and is believed to respond to the nontrivial topology and the corresponding magnon Hall effect \cite{0953-8984-29-3-03LT01,Owerre:2017id}. 
However, a large magnon thermal Hall conductivity can be found even when the scalar spin chirality is very small \cite{PhysRevB.98.094419}. This contradictory 
phenomena suggests that, unlike the cases of ferromagnets, the role of the spin chirality on magnon Hall effects in antiferromagnets still remains puzzling in many aspects.

In the present study, we show that the coplanar {$\vec{q}=0$}  magnetic structure on kagome antiferromagnets  can give a finite thermal Hall effect if the in-plane Dzyaloshinskii-Moriya interaction (DMI) exists. It is obviously that the non-coplanar spins with finite scalar chirality can't be regarded as the single source of topological magnon bands in frustrated antiferromagnets. One needs to investigate the underlying mechanism in more depth. Here we consider magnetically ordered insulating systems, the thermal transport is mainly carried by the quantized spin fluctuations (magnons). Therefore, it is worthwhile to conduct an investigation into the fluctuation of the scalar chirality up to the second order in $\delta\mathbf{S}_{i} = \mathbf{S}_{i} - \la \mathbf{S}_{i} \ra$, i.e.,  $\delta \hat{\chi}_{ijk} = \langle\mathbf{S}_i\rangle\cdot \left( \delta\mathbf{S}_j \times \delta\mathbf{S}_k\right)$, which is more relevant to magnonic transport than does the scalar chirality $\la\hat\chi_{ijk}\ra$. Moreover, even if the scalar chirality $\la\hat\chi_{ijk}\ra = 0$, the fluctuation $\delta \hat{\chi}_{ijk}$ can have a finite value and break dynamically the TRS, leading to topological magnons. Similar to previous studies {\cite{PhysRevLett.104.066403,0953-8984-29-3-03LT01,PhysRevB.98.094419}, 
we examine the DMI term $\mathbf{D}_{jk} \cdot \left(\mathbf{S}_j \times \mathbf{S}_k \right)$, which is rewritten to include an effective ring exchange interaction of three neighbouring spins, $\sim K_{\Phi }/S \left[\mathbf{S}_i\cdot \left(\mathbf{S}_j \times \mathbf{S}_k \right)\right]$ and to introduce explicitly $\delta \hat{\chi}_{ijk}$ to the spin-wave Hamiltonian.    
We show that the component of DM vector ($\mathbf{D}_{jk}$) parallel to the $\la \mathbf{S}_i \ra$ can lead to a finite magnetic flux $K_{\Phi }$ and nontrivial topological effects. The perpendicular component, however, is irrelevant to the thermal Hall effect. Analogous to ferromagnetic systems, the principle (no-go theorem) \cite{PhysRevLett.104.066403} is generalised to rule out the magnon thermal Hall effect in coupled antiferromagnetic systems.

The article is organised as follows. In Sec. \ref{II} we explicitly give out the relation between the DMI and the fluctuation of spin chirality, and emphasize the importance of effective magnetic flux $K_{\Phi} \sim \la \mathbf{D}_{jk}\cdot \mathbf{S}_i \ra$. We show that the (exactly) coplanar spin configuration with an in-plane DMI can give rise to a non-zero $K_{\Phi}$ and induce the nontrivial topology. In Sec.~\ref{model} we introduce the spin model with the in-plane DMI for the kagome antiferromagnets and substantiate the topological nature of magnon bands based on the Holstein-Primakoff method. In Sec.~\ref{IV} the thermal Hall conductivity is calculated with different canting angles. 
Conclusions and discussions are given in Sec.~\ref{V}.

\section{Dzyaloshinskii-Moriya interaction and spin chirality}\label{II}
\begin{figure}
\begin{center}
\vspace{.5cm}
\hspace{-.0cm}\includegraphics[width=1\columnwidth]{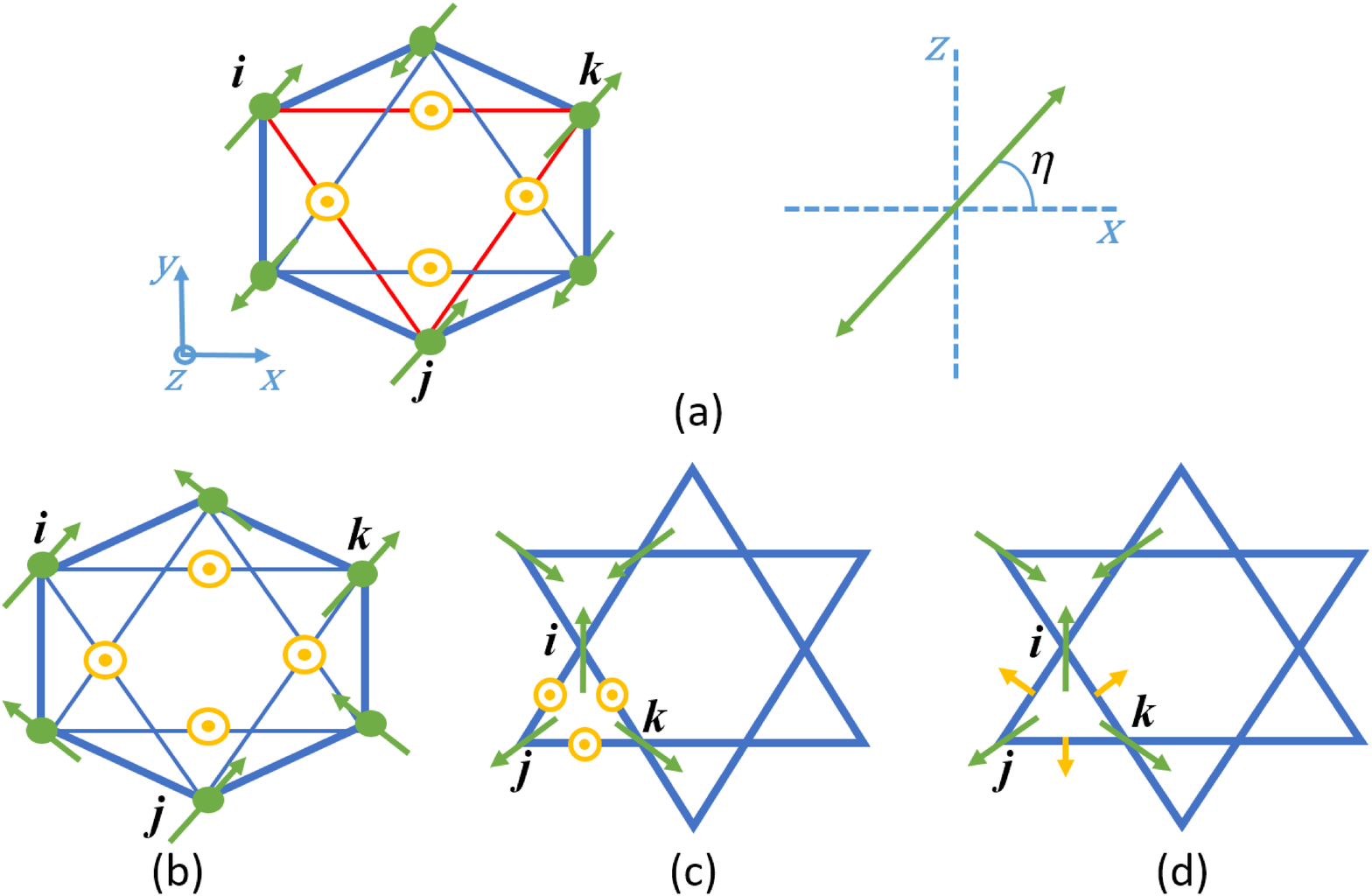}
\vspace{.0cm}
\caption{\label{AFM}(a) The honeycomb lattice with collinear order (left). The DMI (yellow circle) points out-of-plane at the midpoint of the next-nearest-neighbour bonds. The angle between the collinear spin moment and the honeycomb plane is indicated by $\eta$ (right).
(b) The honeycomb lattice with non-collinear order. DMI is same as in (a). 
(c) Coplanar $\mathbf{q}=0$ spin configuration on the
kagome lattice with positive vector chirality. DMI (yellow circle) is perpendicular to the kagome plane at the midpoint of the nearest-neighbour bonds.
(d) Coplanar $\mathbf{q}=0$ spin configuration on the
kagome lattice with positive vector chirality. DMI (yellow arrow) lies in the kagome plane and is perpendicular to the midpoint of the nearest-neighbour bonds.
The green arrows indicate the directions of spins.}
\end{center}
\end{figure}

As pointed out by Katsura \emph{et al.} in Ref. \cite{PhysRevLett.104.066403}, there is a coupling between the scalar chirality and  magnetic fields through the ring exchange process, which, we believe, is the primary interaction source of magnon topological effects in spin systems. In order to get such TRS-broken three-spin interactions in a spin Hamiltonian with the (two-spin) DMI, 
we decompose the DM vector as,
\begin{equation}
\mathbf{D}_{ij} = \mathbf{D}_{ij}^{\perp} + \mathbf{D}_{ij}^{\parallel},
\end{equation}
where $\mathbf{D}_{ij}^{\parallel}$ represents the component whose direction aligns with the third contiguous spin $\mathbf{S}_{k}$ that is determined uniquely through the (cyc. perm.) triangular-bond $\Delta_{ijk}$. $\mathbf{D}_{ij}^{\perp}$ describes the transverse deviation from $\mathbf{S}_{k}$.  If $\mathbf{D}_{ij}^{\parallel} \neq 0$, the DMI can be mapped effectively onto a ring exchange interaction as, 
\begin{equation}
      \mathbf{D}^{\parallel}_{ij}\cdot \left( \mathbf{S}_i \times \mathbf{S}_j \right)=\frac{K_{\Phi}}{S}~\mathbf{S}_k \cdot \left( \mathbf{S}_i \times \mathbf{S}_j\right),
      \label{DZFM}
\end{equation}
where $K_{\Phi} = \mathbf{D}_{ij} \cdot\mathbf{S}_k/S$. The DM vector behaves as an ``orbital magnetic field'' in Eq.~\eqref{DZFM}.
For example, in the collinear ferromagnetic kagome systems, all spin moments are given to be directed normal to the kagome plane and only the out-of-plane DMI is considered $\mathbf{D}_{ij}=\mathbf{D}_z$, one has $K_{\Phi}=D_z \neq 0$ and then finite magnon Hall-type transport \cite{PhysRevLett.104.066403,PhysRevLett.115.106603,PhysRevLett.115.147201}. 
However, the fictitious magnetic flux $K_{\Phi}$ does not always exist due to the dot product between $\mathbf{D}_{ij}$ and $\mathbf{S}_{k}$, especially in the antiferromagnetic systems.

\newcommand{\tabincell}[2]{\begin{tabular}{@{}#1@{}}#2\end{tabular}}  
\begin{center}
\begin{table*}
\caption{\label{Table1}Connection between the spin chirality and the thermal Hall effects in honeycomb and kagome lattices with out-of-plane DMI $\vec{D}_z$.}
\begin{ruledtabular}
\begin{tabular}{llccccc}
  & Magnetic structure & \tabincell{c}{Fictitious magnetic\\ flux $K_{\Phi}$}& Ref.& \tabincell{c}{Thermal Hall\\ coefficient $\kappa_{xy}$} & Mechanism\\ \hline
 Honeycomb & Collinear FM
 & $D_{z}$ &\cite{PhysRevLett.117.227201,0953-8984-28-38-386001}&$\sim D_{z}$ & DMI \\
  &  Collinear AFM
  & 0 &\cite{Owerre:2017kq} & 0& no \\
 & Canted non-collinear AFM & $D_{z}\sin(\eta)$ &\cite{Owerre:2017kq} & $\sim D_{z}\sin(\eta)$ & DMI and Zeeman field\\
 Kagome & Collinear FM &$D_{z}$ & \cite{PhysRevLett.104.066403,2014PhRvB..89m4409M} & $D_{z}$& DMI \\
 & Coplanar $\mathbf{q}=0$ $120^{\circ}$ AFM & 0 & \cite{Owerre:2017id} & 0 & no\\
 & Non-coplanar $120^{\circ}$ AFM & $\left(J+D_{z}\right)\sin(\eta)$ &\cite{Owerre:2017id} &$\sim\left(J+D_{z}\right)\sin(\eta) $ & Non-coplanar chiral spin\\
\end{tabular}
\end{ruledtabular}
\end{table*}\end{center}

As an example, let's revisit the antiferromagnetic honeycomb lattice with out-of-plane DMI, as shown in Fig.~\ref{AFM}(a). 
In the case of collinear spins, the DMI of six bonds can be classified into two classes (corresponding the blue $\bigtriangleup$ and red $\bigtriangledown$ in {Fig.~\ref{AFM}(a)},
\begin{align}\label{DZC}
    \mathbf{D}_{z} \cdot \left( \mathbf{S}_i \times \mathbf{S}_j \right)_{\bigtriangledown}&\Rightarrow
     \frac{D_{z}}{S} \sin(\eta)\langle\mathbf{S}_k\rangle\cdot \left( \delta\mathbf{S}_i \times \delta\mathbf{S}_j\right),\\
     \mathbf{D}_{z} \cdot \left( \mathbf{S}_i \times \mathbf{S}_j \right)_{\bigtriangleup}&\Rightarrow
     -\frac{D_{z}}{S} \sin(\eta)\langle\mathbf{S}_k\rangle\cdot \left( \delta\mathbf{S}_i \times \delta\mathbf{S}_j\right),
\end{align}
where $\eta$ is the angle between the spin plane and the honeycomb plane. Clearly, the scalar chirality fluctuation of upward-pointing triangle $\bigtriangleup$ cancels out the contribution from the downward-pointing triangle $\bigtriangledown$. No matter which value $\eta$ takes, one always has the total $K_{\Phi} = 0$, which implies that the out-of-plane DMI can't serve as the coupling field in the spin-wave Hamiltonian. However, for the case with non-collinear spin configuration as shown in {Fig.~\ref{AFM}(b)}, the cancellation doesn't occur,
\begin{align}\label{DZNC}
    \mathbf{D}_{z} \cdot \left( \mathbf{S}_i \times \mathbf{S}_j \right)_{\bigtriangledown}&\Rightarrow
     \frac{D_{z}}{S} \sin(\eta)\langle\mathbf{S}_k\rangle\cdot \left( \delta\mathbf{S}_i \times \delta\mathbf{S}_j\right),\\
     \mathbf{D}_{z} \cdot \left( \mathbf{S}_i \times \mathbf{S}_j \right)_{\bigtriangleup}&\Rightarrow
     \frac{D_{z}}{S} \sin(\eta)\langle\mathbf{S}_k\rangle\cdot \left( \delta\mathbf{S}_i \times \delta\mathbf{S}_j\right),
\end{align}
where $\eta$ is redefined for the angle between the corresponding spin moment $\langle\mathbf{S}_k\rangle$ and the honeycomb plane. Now $K_{\Phi} \propto  \sin(\eta)$, which deduces that the thermal Hall conductivity is proportional to the canting angle $\eta$, in consistent with the previous results \cite{Owerre:2017kq}.
Similarly, the coupling magnetic flux in kagome antiferromagnets with the out-of-plane DMI and the coplanar $\mathbf{q}=0$ state is zero because of the orthogonality between the DM vector $\vec{D}_{z}$ and adjacent spin $\vec{S}_{k}$ (cf. Fig.~\ref{AFM}(c)).  For comparision, in Table~\ref{Table1}, we list topological properties of different spin configurations on honeycomb/kagome lattices in terms of the fictitious magnetic flux, along with previous studies in the literature. 

Now we turn to the research on the effect of in-plane DMI ($\vec{D}_p$). For the coplanar $\vec{q}=0$ spin configuration on the kagome lattice (as shown in Fig.~\ref{AFM}(d)), the magnetic flux provided by the in-plane DMI can survive since the $\vec{D}_p$ is parallel to $\langle\mathbf{S}_{k}\rangle$,
\begin{align}\label{DP}
    \mathbf{D}_{p} \cdot \left( \mathbf{S}_i \times \mathbf{S}_j \right)&\Rightarrow
     -\frac{D_{p}}{S}\langle\mathbf{S}_k\rangle\cdot \left( \delta\mathbf{S}_i \times \delta\mathbf{S}_j\right).
\end{align}
Noted that the sign in Eq.~\eqref{DP} is opposite to the contribution given by the non-zero scalar chirality \cite{Owerre:2017id}. Consequently the coupling fields provided respectively by the in-plane DMI and the scalar chirality will weaken each other (in the Sec.~\ref{IV} we will discuss this cancellation in more detail).
From the symmetry point of view, 
the so-called (non-coplanar) umbrella spins with non-zero scalar chirality $\langle \chi_{ijk}\rangle \neq 0$ and a weak ferromagnetic moment breaks the TRS statically \cite{PhysRevB.66.014422,PhysRevLett.96.247201,PhysRevB.98.094419}.  However, the fluctuation of scalar chirality $\delta \chi_{ijk}$ can be finite from the mean-field argument even though $\langle \chi_{ijk} \rangle =0$ in the case of coplanar magnetic orders, Eq.~\eqref{DP} suggests a dynamically TRS-broken interaction in the spin-wave Hamiltonian. A finite topological magnon Hall effect is thus expected in the coplanar spin configuration with the in-plane DMI. 
\section{Spin model and topological magnons}\label{model}
\begin{figure}[b]
\begin{center}
\vspace{.5cm}
\hspace{-.0cm}\includegraphics[width=\columnwidth]{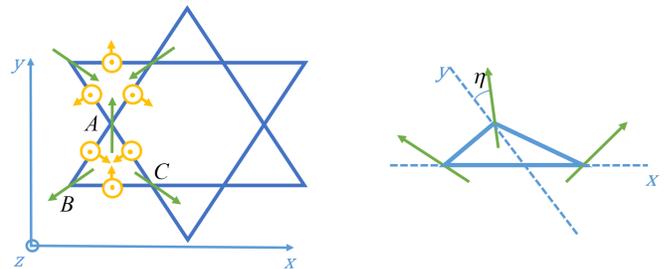}
\vspace{.0cm}
\caption{\label{kagome}Non-coplanar spin configuration on the kagome lattice with positive vector chirality. All the spins have a weak $z$ component resulting in weak ferromagnetic angle $\eta$. The yellow arrows and circles in the middle point between the magnetic sites represent the in-plane DMI and out-plane DMI, respectively.}
\end{center}
\end{figure}

Let's consider the spin model on the frustrated kagome lattice, in which the DMI comes naturally because of lacking of an inversion center \cite{DZYALOSHINSKY1958241,PhysRev.120.91},
\begin{align}\label{H1}
 H&= \sum_{i,j}  J_{ij}{\bf S}_{i}\cdot{\bf S}_{j} -\sum_{\la i, j\ra} \mathbf{D}_{ij}\cdot{\bf S}_{i}\times{\bf S}_{j}-B\hat{\bf z}\cdot\sum_i \mathbf{S}_{i},
\end{align}
where the first term contains the nearest-neighbour (NN) and the next-nearest-neighbour (NNN) antiferromagnetic interactions $J_1$ and $ J_2$, respectively. $\mathbf D_{ij}=(0, D_p, D_z)$ with $D_p$ and $D_z$ being the in-plane and the out-of-plane DMI for the bond ($ij$), respectively (cf. Fig.~\ref{kagome}). 
$\vec{B}$ is the normal magnetic field in units of $g\mu_B$. The influence of the DMI in the frustrated kagome lattice has been studied extensively. For the spin-$1/2$ case, both exact diagonalization method and the Schwinger bosons mean field theory have predicted a quantum transition between the quantum spin liquid with a small $|D_z|$ and the $\vec{q}=0$ N\'eel state as $|D_z|>D^{c}_{z}\approx0.1J_1$ \cite{PhysRevB.78.140405,PhysRevB.81.064428,PhysRevB.81.144432,PhysRevLett.98.207204,PhysRevB.88.224413}. For higher spins, an ordered $120^{\circ}$ magnetic configuration is favorite in energy \cite{PhysRevB.66.014422}. The {out-of-plane} DMI $D_z$ stabilizes the $120^{\circ}$ coplanar $\vec{q}=0$ spin structure \cite{PhysRevB.66.014422}. Even if $D_z=0$, the coplanar structure can be equally stabilized by the NNN coupling $J_2$ \cite{PhysRevB.45.2899}. Depending on the sign of $D_z$, two vector chiralities can be allowed in the kagome lattice. As shown in Fig.~\ref{kagome}, we choose that $D_z>0$ corresponds to the positive vector chirality in the Ref. \onlinecite{PhysRevB.66.014422}. The in-plane DMI ${D}_{p}$ breaks mirror reflection symmetry with respect to the kagome plane and the global spin rotation symmetry \cite{PhysRevB.66.014422,PhysRevLett.96.247201}. It prefers umbrella spins. However, the vector chirality that stems from the in-plane component $D_{p}$ is the same as the one selected by $D_z>0$. 
By taking spin $\vec{S}_{i}$ as a classical vector, the canting angle of  the $120^{\circ}$ spin structure up to the first order can be expressed by (see the Appendix~\ref{A} for details),
\begin{align}\label{canting}
      \eta	= \frac{B/S + 2\sqrt{3} D_p }{6(J_1+J_2)+ 2\sqrt{3} D_z}.	
\end{align}
For a small $\eta$, Eq.~\eqref{canting} is a good approximation. It is obvious that one can eliminate the canting angle to recover the coplanar spin structure by balancing the in-plane DMI with applied external magnetic field $B=-2\sqrt{3}SD_p$. Consequently, we have a very good opportunity to check immediately whether the dynamic $\delta \chi_{ijk}$ (but with $\chi_{ijk} =0$) can result in topological magnon bands or not.  

The spin wave {excitations} of the $120^{\circ}$ coplanar $\vec{q}=0$ spin structure with balanced $D_{p}$ and $B$ {is} investigated by  Holstein-Primakoff bosons (see Appendix~\ref{A}). Figure~\ref{bulkab} shows the bulk magnon bands 
along the high-symmetry points of the Brillouin zone with ${\bf\Gamma}=(0,0)$, ${\bf K}=(2\pi/3,0)$, and ${\bf M}=(\pi/2,\pi/2\sqrt{3})$. The parameters are from the ideal kagome antiferromagnets such as iron jarosite KFe$_3$(OH)$_6$(SO$_4$)$_2$ \cite{PhysRevLett.96.247201} and veseignite BaCu$_3$V$_2$O$_8$(OH)$_2$ \cite{PhysRevB.88.144419}. There are two evident features in the figures: (i) One Goldstone mode at the ${\bf\Gamma}$ point due to regaining of the SO(2) rotation symmetry by adjusting magnetic field $B$ to compensate the canting angle induced by the in-plane $D_p$. 
(ii) Finite gap appears in the rest of the Brillouin zone. It should be noted that these gaps do not go to zero in the thermodynamic limit even if they look very small. 
\begin{figure*}
\includegraphics[width=1\textwidth]{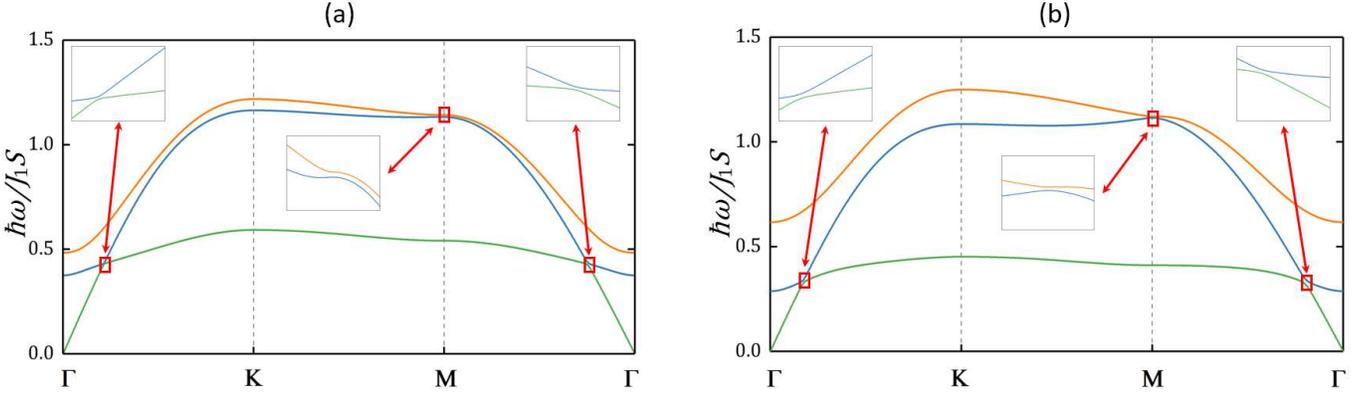}
\caption{\label{bulkab}The magnon bulk bands of the coplanar spin configuration with the canting angular $\eta=0$. (a) $D_z/J_1=0.062$, $D_p/J_1=0.062$, $J_2/J_1=0.035$, and $B/J_1=-0.537$ with $J_1=3.18$ meV corresponding to iron jarosite KFe$_3$(OH)$_6$(SO$_4$)$_2$. (b) $D_z/J_1=0.07$, $D_p/J_1=0.19$, $J_2/J_1=0$, and $B/J_1=-0.329$ with $J_1=4.6$ meV corresponding to veseignite BaCu$_3$V$_2$O$_8$(OH)$_2$. The insets magnify the gaps of magnon bands indicated by red squares.}
\end{figure*}

In general, the dynamic $\delta \chi_{ijk}$ is composed of three main contributions,
\begin{align}\label{flux}
      H^{\Phi}	&= H^{\Phi}_{J_1}+H^{\Phi}_{D_z}+H^{\Phi}_{D_p},\\
      H^{\Phi}_{J_1} &= -J_1\frac{\sqrt{3}}{2}\sin{\eta}\hat{\bf z}\cdot\lb \mathbf{S}^{\prime}_i \times \mathbf{S}^{\prime}_j \rb,\\
      H^{\Phi}_{D_z} &= D_z\frac{1}{2}\sin{\eta}\hat{\bf z}\cdot\lb  \mathbf{S}^{\prime}_i \times \mathbf{S}^{\prime}_j \rb,\\
      H^{\Phi}_{D_p} &= D_p\frac{1}{2}\cos{\eta}\hat{\bf z}\cdot\lb  \mathbf{S}^{\prime}_i \times \mathbf{S}^{\prime}_j \rb.
\end{align}
The corresponding fictitious magnetic flux reads $K_{\Phi} \sim \lb -J_1\sqrt{3}\sin{\eta} +D_z\sin{\eta} + D_p\cos{\eta}\rb$, which is closely coupled to the $z$-polarized (magnon) spin current \cite{doi:10.7566/JPSJ.86.011007}. Evidently, the dynamic scalar chirality  $\delta \chi_{ijk}$ survives even at  $\eta = 0$. Now the in-plane DMI becomes an only source of topological spin excitations in the $\vec{q}=0$ coplanar magnetic ordering, which possesses different mechanism from the one with non-coplanar magnetic configuration induced by external magnetic fields \cite{0953-8984-29-3-03LT01,Owerre:2017id}. Furthermore, the $H^{\Phi}_{D_p}$ won't disappear immediately even if the coplanar spins are deformed into the umbrella configuration. The influence of the in-plane DMI on the {magnon bands} of non-coplanar configuration will be discussed in Sec.~\ref{IV}.

In order to confirm the topological nature of magnons, the chiral edge states are demonstrated in the Fig.~\ref{Edge} using the strip geometry with open boundary conditions along $y$ direction and infinite along $x$ direction. We clearly see crossed chiral edge modes between the middle and lower bands from Fig.~\ref{Edge} (a) and (b). This forcefully supports the existence of topological magnons driven by only the in-plane DMI. 
Another evidentiary quality of topological magnons is the non-zero Chern number of the bulk bands. Unlike electron systems, the Chern number of bosonic systems is not well-defined because of the lack of the Fermi surface and an evenly filled band. However, we can still calculate the Chern number for the $n$-th bulk band mathematically as
\begin{align}\label{Chern}
      C_{n} = \frac{1}{2\pi}\int_{BZ}\Omega_{n{\bf k}}~d^2{\bf k},
\end{align}
where $\Omega_{n{\bf k}}$ is the Berry curvature for bosonic Bogliubov-de Gennes systems. Noted that the formula of $\Omega_{n{\bf k}}$ has to be modified accordingly (See Appendix~\ref{B}) \cite{PhysRevB.87.174427}. The Chern numbers in Fig.~\ref{bulkab} (a) and (b) are given as $(+3,-1,-2)$ going from the lower, middle, and upper bands. When the $D_p$ drops to zero, both the gapless chiral edge modes and the Chern numbers will disappear. This is insofar conclusive: the non-trivial topological spin excitations of coplanar magnetic structures must be purely tied to the dynamic spin chirality given by the in-plane DMI (cf. also Eq.~\eqref{DP}). For comparison, the edge modes under the influence of $D_p$ with the large canting angle $\sin\eta =0.4$ are also shown in Fig.~\ref{Edge} (c), which is almost as same as the case without $D_p$ in Ref.~\cite{Owerre:2017id}. This seemingly suggests that the $D_p$ is negligible for the topological property of the non-coplanar spin configuration as in the previous studies \cite{Owerre:2017id} when the canting angle $\eta$ is quite large. However, one should be careful about this conclusion because the Chern number is the feature of the entire Brillouin zone. Actually, the calculations show that the Chern number $C_{n}=(-1,+2,-1)$ for magnon bands with $D_p$ is different from $C_{n}=(-1,+4,-3)$ for the bands in the absence of $D_p$. The change of Chern numbers stems from the symmetry breaking. As we mentioned, $D_p$ breaks the SO(2) spin rotation symmetry and tends to give rise to weak out-of-plane ferromagnetic component. As a result, the zero-energy Goldstone mode at the $\vec\Gamma$ point is lifted in the non-coplanar spins by the in-plane DMI. A new gap at the ${\bf\Gamma}$ point increases with the external field $B$. The ``monopole'' at the ${\bf\Gamma}$-${\bf K}$ line is dissolved as the magnetic field $B$ reaches the threshold value. In Figure~\ref{Berry}, we demonstrate the substantial change happening in the magnon band structures and Berry curvature of the non-coplanar spins with/without $D_{p}$.
\begin{figure}
\begin{center}
\includegraphics[width=1\columnwidth]{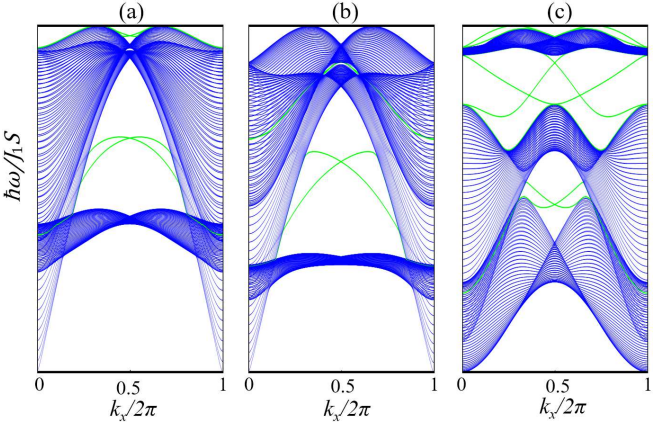}
\caption{\label{Edge}Edge states in kagome antiferromagnets with in-plane DMI. (a) $\eta=0$, parameters corresponding to iron jarosite KFe$_3$(OH)$_6$(SO$_4$)$_2$ (same as in {Fig.~\ref{bulkab}(a)}). (b) $\eta=0$, parameters corresponding to veseignite BaCu$_3$V$_2$O$_8$(OH)$_2$  (same as in Fig.~\ref{bulkab}(b)). (c) $\sin\eta=0.4$, $D_z/J_1=0.06$, $D_p/J_1=0.06$, and $J_2/J_1=0.03$ corresponding to the parameters in Ref.~\cite{Owerre:2017id} with in-plane DMI appended. The blue lines are the gapped bulk bands and the green lines are the gapless edge modes.}
\end{center}
\end{figure}

\begin{figure}
\begin{center}
\includegraphics[width=1\columnwidth]{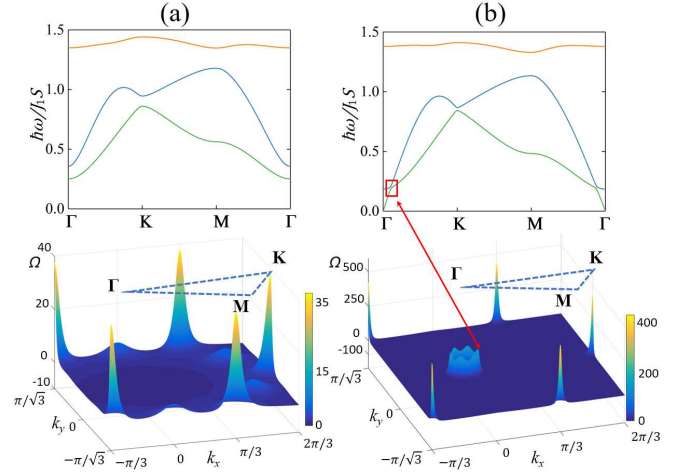}
\caption{\label{Berry}The magnon bulk bands (upper) and the corresponding Berry curvature of the middle band (bottom). (a) $D_p/J_1=0.06$. (b) $D_p/J_1=0$. Other parameters are the same as in Fig.~\ref{Edge}(c). The ``monopole'' at the ${\bf\Gamma}$-${\bf K}$ line is indicated by the red square.}
\end{center}
\end{figure}

\section{Topological thermal Hall effect}\label{IV}
The thermal Hall effect is a powerful probe  to unveil the topological nature of low-energy quasipaticle excitations in magnetic systems. The magnon thermal Hall conductivity $\kappa_{xy}$ is closely connected to the Berry curvature \cite{PhysRevB.89.054420},
\begin{eqnarray}
  \kappa_{xy} &=& - \frac{k_B^2 T}{\hbar V}
  \sum_{{\bf k}\in BZ}\sum^{N}_{ n=1} \left \{
    c_2\left[g\lb \varepsilon_{n{\bf k}}\rb \right] - \frac{\pi^2}{3}
\right \} \Omega_{n {\bf k}}.
\end{eqnarray}
Here, $g\lb \varepsilon_{n{\bf k}}\rb$ is the Bose distribution function $g\lb \varepsilon_{n{\bf k}}\rb = 1/\left[\exp\lb \varepsilon_{n{\bf k}}/k_B T \rb -1\right]$, $c_2(x)$ is defined as $c_2(x) =
(1+x)\left( \ln \frac{1+x}{x} \right)^2 - \left(\ln x\right)^2 - 2
{\rm Li}_2(-x)$. 
The (low) temperature dependence of $\kappa_{xy}$ of the coplanar spin configuration is plotted in Figure~\ref{ka} with the parameters of KFe$_3$(OH)$_6$(SO$_4$)$_2$ and BaCu$_3$V$_2$O$_8$(OH)$_2$, respectively. Both showing monotonically increasing behaviour, very similar to the previous temperature dependence of the non-coplanar configuration caused by $D_{p}$ in Ref.~\cite{PhysRevB.98.094419}. 
That means that the in-plane DMI dominates the Hall-type spin transport in the system even if there is a small canting angle. However, while the canting angle increases {with $D_{p}$ (cf. Eq.~\eqref{canting})}, the role of the in-plane DMI on $\kappa_{xy}$ of the non-coplanar spin configures is complex as shown in the following. 

\begin{figure}[b]
\begin{center}
\vspace{.5cm}
\hspace{-.0cm}\includegraphics[width=0.9\columnwidth]{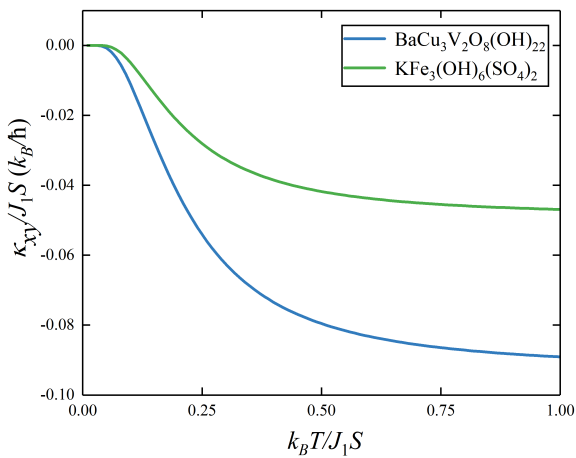}
\vspace{.0cm}
\caption{\label{ka} Low temperature dependence of thermal Hall conductivity $\kappa_{xy}$ with the canting angle $\eta=0$.}
\end{center}
\end{figure}

\subsection{Small canting angle}
\begin{figure}
\begin{center}
\vspace{.5cm}
\hspace{-.0cm}\includegraphics[width=0.8\columnwidth]{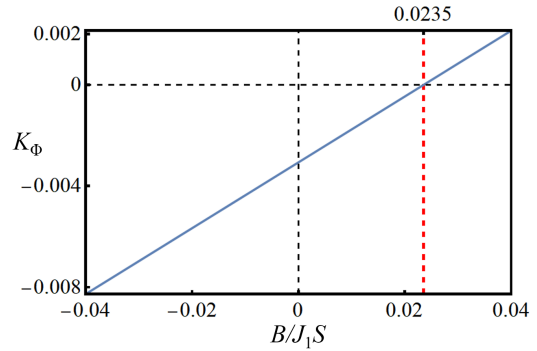}
\vspace{.0cm}
\caption{\label{TP}The flux $K_{\Phi}$ as a function of applied magnetic field $B$ based on Eq.~\eqref{ETP} with the parameters of KFe$_3$(OH)$_6$(SO$_4$)$_2$.}
\end{center}
\end{figure}

As given in the Eq.~\eqref{flux}, the fictitious magnetic flux $K_{\phi}$ is composed of three main contributions as $\eta \neq 0$. 
When the canting angle $\eta$ is smaller {than about $0.1$}, we have $H^{\Phi}_{D_p} \sim D_{p}\lb 1-\eta^2\rb$, $H^{\Phi}_{J_1} \sim J_{1} \eta$, and $H^{\Phi}_{D_z} \sim D_{z}\eta$ (which is one order less than $H^{\Phi}_{J_1}$ given that the out-of-plane $D_{z}/J_{1} < 0.1$). The topological properties are now determined by both the non-zero scalar chirality $\chi_{ijk}$ (mainly through $H^{\Phi}_{J_1}$) and the dynamic fluctuation $\delta\chi_{ijk}$ (principally from $H^{\Phi}_{D_p}$). Moreover, when the canting angle is very small ($\eta \approx 0$), the dynamic $\delta\chi_{ijk}$ becomes the dominant mechanism of topological magnon bands. 
So it is not strange that one can still have a large thermal Hall effect even the scalar chirality $  \chi_{ijk}  \approx 0$. 
On the other hand, to exclude a phononic contribution from the thermal transport, the sign changes in $\kappa_{xy}$ induced by the Chern number ($K_{\Phi}$) sign alternation is incisive when varying the external magnetic field. Such the phase transition point can be estimated based on the Eq.~\eqref{canting} and Eq.~\eqref{flux}, 
\begin{align}\label{ETP}
K_{\Phi} \sim &J_1\sqrt{3}\eta -D_z\eta - D_p \lb 1-\eta^2\rb\\
\nonumber &=\alpha\cdot B/S+\beta+\mathcal{O}(B^2),\\
\nonumber\alpha=&\frac{\sqrt{3}J_1-D_z}{6\lb J_1+J_2 \rb+2\sqrt{3}D_z};\\
\nonumber\beta=&\lb\frac{6J_1-2\sqrt{3}D_z}{6\lb J_1+J_2 \rb+2\sqrt{3}D_z}-1\rb D_p.
\end{align}
Figure~\ref{TP} indicates the critical value of magnetic field, $B/J_1S\sim 0.0235$ (i.e., $B/J_1\sim 0.0588$ with $S=\frac{5}{2}$) with the parameters of KFe$_3$(OH)$_6$(SO$_4$)$_2$, which is in consistent with the numerical result ($B/J_1\sim 0.06$) \cite{PhysRevB.98.094419}. Given that $J_{1} = 3.18$ meV, $\kappa_{xy}$ undergoes sign changes at $B \sim1.6$ Tesla, which can in turn be used for valuing the strength of the in-plane DMI of the frustrated kagome antiferromagnets.

\subsection{Large canting angle}
For the non-coplanar umbrella spins with large canting angle ($\cos \eta \ll 1$), Eq.~\eqref{flux} indicates that the contribution of the in-plane DMI $H^{\phi}_{D_p}$ is tiny compared to the $H^{\phi}_{J_1}$ and $H^{\Phi}_{D_z}$. It seems that the effect of  $H^{\phi}_{D_p}$ on the magnon topological properties could be negligible.  
However, it is not true from the point view of spin symmetry. As shown in Fig.~\ref{Berry}, the in-plane $D_p$ breaks the global spin rotation symmetry and results in a finite gap at the $\Gamma$ point. We have then fully gapped magnon bands in the entire Brillouin zone, which provides an energy barrier for thermal excitations. Thus, it is anticipated that the thermal Hall conductivity with the in-plane DMI will be smaller than the value without $D_p$,   as indicated by the $c_2$ function as well \cite{PhysRevB.89.054420}.  
 Figure~\ref{kb} displays the comparison of $\kappa_{xy}$ with/without $D_p$ under the large canting angle, respectively. As expected, the presence of in-plane DMI blocks the excitations and suppresses the thermal Hall effect. 
 
\begin{figure}
\begin{center}
\vspace{.5cm}
\hspace{-.0cm}\includegraphics[width=0.9\columnwidth]{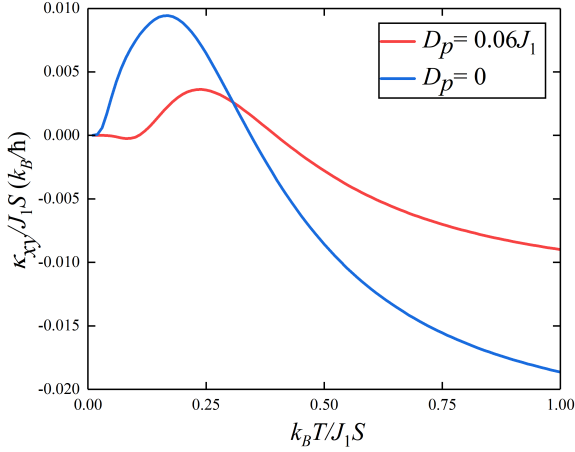}
\vspace{.0cm}
\caption{\label{kb} Low temperature dependence of thermal Hall conductivity $\kappa_{xy}$ with $\sin\eta=0.4$,
$D_z/J_1=0.06$, and $J_2/J_{1}=0.03$.}
\end{center}
\end{figure}
\section{Conclusions}\label{V}
In summary, we have shown that the fluctuation of scalar spin chirality $\delta \chi_{ijk}$ is directly related to the magnonic transport in spin systems.  For the magnetically ordered system with nonzero $\delta \chi_{ijk}$ and fictitious magnetic flux $K_{\Phi}$ in terms of the DMI, the time-reversal symmetry is dynamically broken and topologically nontrivial bands of magnons can be expected.  Based on these considerations and detail analysis for the frustrated kagome antiferromagnets with the in-plane DMI, 
it was found that the in-plane DMI can give rise to topological magnon bands with chiral edge modes and non-zero Chern number even in the case of coplanar $\vec{q} =0$ N\'eel state. The proposed model has been applied to real antiferromagnets to discuss the effects of in-plane DMI on the magnon thermal Hall responses and resolve the contradictory of large thermal Hall conductivities but with tiny spin chirality.  For the non-coplanar spins with large canting angle $\eta$, the gapped magnons induced by the in-plane DMI is considerably suppressed the thermal excitations and reduces the thermal Hall conductivity $\kappa_{xy}$. In this regard, the present study generalises the  \emph{No-go} theorem for the antiferromagnetically coupled systems and clarifies the mechanism of  nontrivial topology in magnon systems. 
\section{Acknowledgments}
This work is supported by the National Natural Science Foundation of China (No. 11474138 and No. 11834005), the German Research Foundation (No. SFB 762), the Program for Changjiang Scholars and Innovative Research Team in University (No. IRT-16R35), the Fundamental Research Funds for the Central Universities, Ministry of Science and Technology of China through grants CN-SK-8-4, the Slovak Academy of Sciences (Grant No. 2/0059/17) and the Slovak Research and Development Agency (APVV SK-CN-2017-0004).

\onecolumngrid
\appendix
\section{The magnon tight binding Hamiltonian in the
Holstein-Primakoff mean-field theory}\label{A}
To study the fluctuations of the spin wave, we first need to determine the classical groundstate of the system. In the classical limit, the SU(2) spin operators can be replaced by the classical SO(3) vectors, $\mathbf{S}=S\lb\cos\theta\cos\eta,~\sin\theta\cos\eta, ~\sin\eta \rb$, where $\eta$ is the canting angle. The oriented
angles of three sub-lattices are $\theta_A = \pi/2$, $\theta_B = 7\pi/6$, $\theta_C =-\pi/6$ for the $120^{\circ}$ magnetic configuration. Thus, the classical energy reads \cite{PhysRevB.66.014422} 
\begin{align}\label{Energy}
\frac{E^{cl}}{\mathcal{N}S^2}=&\lb J_1+J_2\rb\lb1-3\cos2\eta\rb/2-\sqrt{3}D_z\cos^2\eta-\sqrt{3}D_p\sin2\eta-B\sin\eta/S.
\end{align}
By minimizing the energy, the canting angle $\eta$ is determined by,
\begin{align}\label{extreme}
0= \frac{\partial}{\partial \eta}\lb\frac{E^{cl}}{\mathcal{N}S^2}\rb=&3\lb J_1+J_2\rb\sin2\eta+\sqrt{3}D_z\sin2\eta-2\sqrt{3}D_p\cos2\eta-B\cos\eta/S.
\end{align}
Up to the first order, $\eta$ is given by
\begin{align}\label{Acanting}
 \eta	\approx \frac{B/S + 2\sqrt{3} D_p }{6(J_1+J_2)+ 2\sqrt{3} D_z}.
\end{align}
For the magnonic properties, it is necessary to perform a local coordinate transformation for getting the spin-wave excitation above the groudstate,
\begin{align}\label{lct}
  \nonumber \mathbf{S}_A =& \lb S^{\prime x}_A, \cos\eta S^{\prime z}_A+\sin\eta S^{\prime y}_A, -\cos\eta S^{\prime y}_A+\sin\eta S^{\prime z}_A\rb,\\
  \nonumber\mathbf{S}_B =& \lb  -\frac{1}{2}S^{\prime x}_B-\frac{\sqrt{3}}{2}\lb\cos\eta S^{\prime z}_B+\sin\eta S^{\prime y}_B\rb,\frac{\sqrt{3}}{2}S^{\prime x}_B-\frac{1}{2}\lb\cos\eta S^{\prime z}_B+\sin\eta S^{\prime y}_B\rb, -\cos\eta S^{\prime y}_B+\sin\eta S^{\prime z}_B\rb,\\
  \mathbf{S}_C =& \lb  -\frac{1}{2}S^{\prime x}_C+\frac{\sqrt{3}}{2}\lb\cos\eta S^{\prime z}_C+\sin\eta S^{\prime y}_C\rb,-\frac{\sqrt{3}}{2}S^{\prime x}_C-\frac{1}{2}\lb\cos\eta S^{\prime z}_C+\sin\eta S^{\prime y}_C\rb, -\cos\eta S^{\prime y}_C+\sin\eta S^{\prime z}_C\rb,
\end{align}
where $\mathbf{S}^{\prime}_i$ is the local spin operator, $\lb A,~B,~C\rb$ are three sublattice sites as depicted in Fig.~\ref{kagome}. Then the Hamiltonian Eq.~\eqref{H1} can be rewritten as
\begin{align}\label{H2}
      &H	= H_{J_1}+H_{J_2}+H_{D_z}+H_{D_p}+H_z,\\
      &H_{J_1}=J_1\sum_{\langle ij\rangle}\left[\cos\frac{2}{3}\pi~\mathbf{S}^{\prime}_i\cdot\mathbf{S}^{\prime}_j-\sin\frac{2}{3}\pi\sin\eta\hat{\bf z}\cdot\lb \mathbf{S}^{\prime}_i \times \mathbf{S}^{\prime}_j \rb+2\sin^2\frac{1}{3}\pi\lb\cos^2\eta S^{\prime y}_iS^{\prime y}_j+\sin^2\eta S^{\prime z}_iS^{\prime z}_j\rb\right],\\
      &H_{J_2}=J_2\sum_{\langle\langle ij\rangle\rangle}\left[\cos\frac{2}{3}\pi~\mathbf{S}^{\prime}_i\cdot\mathbf{S}^{\prime}_j-\sin\frac{2}{3}\pi\sin\eta\hat{\bf z}\cdot\lb \mathbf{S}^{\prime}_i \times \mathbf{S}^{\prime}_j \rb+2\sin^2\frac{1}{3}\pi\lb\cos^2\eta S^{\prime y}_iS^{\prime y}_j+\sin^2\eta S^{\prime z}_iS^{\prime z}_j\rb\right],\\
      &H_{D_z}=-D_z\sum_{\langle ij\rangle}\left[\sin\frac{2}{3}\pi\lb S^{\prime x}_iS^{\prime x}_j+\sin^2\eta S^{\prime y}_iS^{\prime y}_j+\cos^2\eta S^{\prime z}_iS^{\prime z}_j\rb+\cos\frac{2}{3}\pi\sin\eta\hat{\bf z}\cdot\lb \mathbf{S}^{\prime}_i \times \mathbf{S}^{\prime}_j \rb\right],\\
      &H_{D_p}=-D_p\sum_{\langle ij\rangle}\left[\sin\frac{2}{3}\pi\sin2\eta\lb S^{\prime z}_iS^{\prime z}_j- S^{\prime y}_iS^{\prime y}_j\rb+\cos\frac{2}{3}\pi\cos\eta\hat{\bf z}\cdot\lb \mathbf{S}^{\prime}_i \times \mathbf{S}^{\prime}_j \rb\right],\\
      &H_z=-B\sin\eta\sum_{i}S^{\prime z}_{i}.
\end{align}
Obviously, several terms that can be rewritten as the fictitious magnetic field coupling with spin chirality provide anomalous velocity of the magnons, they are
\begin{align}\label{Aflux}
 H^{\Phi} = \lb -J_1\frac{\sqrt{3}}{2}\sin{\eta} +\frac{D_z}{2}\sin{\eta} + \frac{D_p}{2}\cos{\eta}\rb\hat{\bf z}\cdot\lb  \mathbf{S}^{\prime}_i \times \mathbf{S}^{\prime}_j \rb.
\end{align}
We have neglected the flux term coming from $H_{J_2}$. Because the  magnitude of $J_2\sin\eta$ is far less than $J_{1}\sin\eta$ and $D_p\cos\eta$  under the condition of small canting angle $\eta$. 

Following the Holstein-Primakoff approach, the local spin operators are expressed by the bosonic annihilation and creation operators as $S_i^{\prime x}=\sqrt{S/2}(b_i^\dg+b_i)$, $S_i^{\prime y}=i\sqrt{S/2}(b_i^\dg-b_i,)$, and $S_i^{\prime z}=S-b_i^\dg b_i$ \cite{PhysRev.58.1098}. The magnon tight binding Hamiltonian becomes
 \begin{align}\label{SW}
 H_{SW}=& S\sum_{\mathbf{k},\alpha,\beta; 1,2}2\lb M_{\alpha\beta}^0\delta_{\alpha\beta} +M_{\alpha\beta; 1,2}\rb b_{\mathbf{k} \alpha}^\dagger b_{\mathbf{k} \beta}+M_{\alpha\beta; 1,2}^{\prime} \lb b_{\mathbf{k} \alpha}^\dagger b_{-\mathbf{k} \beta}^\dagger +b_{\mathbf{k} \alpha} b_{-\mathbf{k} \beta}\rb,
\end{align}
where $\alpha,\beta=A,B,C$ and the coefficient matrixs are expressed by
$\mathbf{M}^0=M^0\mathbf{I}_{3\times 3}$ with $M^0=\lb J_1+J_2\rb\lb 1-3\sin^2\eta\rb+D_z\sqrt{3}\cos^2\eta+D_p\sqrt{3}\sin2\eta+B\sin\eta/2S$.

\begin{align}
&\mathbf{M}_{1,2}= M_{1,2}\begin{pmatrix}
  0& \gamma_{AB}^{1,2}e^{-i\phi_{1,2}}&\gamma_{CA}^{1,2}e^{i\phi_{1,2}} \\
\gamma_{AB}^{*1,2} e^{i\phi_{1,2}}& 0&\gamma_{BC}^{1,2}e^{-i\phi_{1,2}}\\
\gamma_{CA}^{*1,2} e^{-i\phi_{1,2}}& \gamma_{BC}^{*1,2} e^{i\phi_{1,2}} & 0 \\
 \end{pmatrix} ~~\text{and}
 &\mathbf{M}_{1,2}^\prime=M_{1,2}^\prime\begin{pmatrix}
  0& \gamma_{AB}^{1,2}&\gamma_{CA}^{1,2} \\
\gamma_{AB}^{*1,2} & 0&\gamma_{BC}^{1,2}\\
\gamma_{CA}^{*1,2} & \gamma_{BC}^{*1,2} & 0 \\
 \end{pmatrix},
\end{align}
where $\gamma_{AB}^1=\cos k_1,~\gamma_{BC}^1=\cos k_2,~\gamma_{CA}^1=\cos k_3;~\gamma_{AB}^2=\cos p_1,~ \gamma_{BC}^2=\cos p_2,~\gamma_{CA}^2=\cos p_3$ and $p_i=\mathbf{p}\cdot\mathbf{e}_i^\prime,~\mathbf{e}_1^\prime=\mathbf{e}_3-\mathbf{e}_2,~\mathbf{e}_2^\prime=\mathbf{e}_1-\mathbf{e}_3,~\mathbf {e}_3^\prime=\mathbf{e}_2-\mathbf{e}_1$. The normalized fluxes are given by  $\phi_{1,2}=\tan^{-1}\lb M_{1,2}^{im}/M_{1,2}^{re}\rb$, and $M_{1,2}=\sqrt{(M_{1,2}^{re})^2+(M_{1,2}^{im})^2}$, where
\begin{align}
  &M^{re}_1= \frac{1}{4}J_1\lb3\cos^2\eta-2\rb-\frac{\sqrt{3}}{4}D_z\lb 1+\sin^2\eta\rb+\frac{\sqrt{3}}{4}D_p\sin2\eta,\\
  &M^{im}_1=\lb-\frac{\sqrt{3}}{2}J_1+\frac{1}{2}D_z\rb\sin\eta+\frac{1}{2}D_p\cos\eta,\\
  &M_{1}^{\prime}=\frac{1}{4}\lb3J_1\cos^2\eta+\sqrt{3}D_z\cos^2\eta+\sqrt{3}D_p\sin2\eta\rb,\\
  &M^{re}_{2}=\frac{1}{4}J_2\lb3\cos^2\eta-2\rb,\\
  &M^{im}_{2}=-\frac{\sqrt{3}}{2}J_2\sin\eta.
 \end{align}
It needs to be emphasized that the flux does not vanish even if both $\eta$ and $D_z$ are equal to zero. This means that the only $D_p$ with coplanar spin structure can break the TRS as well. 

By introducing Nambu spinor
 \begin{align}\label{spinor}
\left[\beta^{\dg}_{\mathbf{k}} ~~\beta_{-\mathbf{k}} \right]\equiv \left[ b_{\mathbf{k} A}^{\dg},~ b_{\mathbf{k} B}^{\dg},~ b_{\mathbf{k} C}^{\dg},~ b_{-\mathbf{k} A},~ b_{-\mathbf{k} B},~b_{-\mathbf{k} C} \right],
 \end{align}
The Hamiltonian can be written as
 \begin{align}\label{Bdg}
H_{SW}=E_0+ S\sum_{\mathbf{k}}
\left[\begin{array}{cc}
{\bm \beta}^{\dag}_{\mathbf{k}} &
{\bm \beta}_{-\mathbf{k}}\\
\end{array}\right] \cdot \mathbf{H}_{\mathbf{k}}
\cdot \left[\begin{array}{c}
{\bm \beta}_{\mathbf{k}} \\
{\bm \beta}^{\dag}_{-\mathbf{k}} \\
\end{array}\right].
 \end{align}

\section{Berry curvature for bosonic Bogliubov-de
Gennes systems}\label{B}
Although the Chern number can be still defined as Eq.~\eqref{Chern} for the bosonic BdG system, the formula of Berry curvature is different from the fermion system. A general formalism is developed for bosonic BdG systems by Shindou \cite{PhysRevB.87.174427}. Unfortunately, this method contains derivatives of the eigenstates that can not be used directly for numerical calculations. Here the gauge-independent formula of Berry curvature are provided for bosonic BdG systems.
We consider a quadratic form of generic bosonic Hamiltonian as given by Eq.~\eqref{Bdg}. Such a bosonic BdG Hamiltonian can be diagonalized by the Bogliubov transformation by using the para-unitary transformation $\mathcal{T}_{\mathbf{k}}$ instead of the unitary transformation,
\begin{eqnarray}\label{para1}
{\mathcal T}^{\dg}_{\mathbf{k}} \!\ \textbf{H}_{\mathbf{k}}
\!\ {\mathcal T}_{\mathbf{k}} = \left[\begin{array}{cc}
{E}_{\mathbf{k}} & \\
& {E}_{-\mathbf{k}} \\
\end{array}\right]
\end{eqnarray}
with $[\gamma^{\dg}_{\mathbf{k}}, \gamma_{-\mathbf{k}}] \!\ { \mathcal{T}}^{\dag}_{\mathbf{k}} =
[\beta^{\dg}_{\mathbf{k}}, \beta_{-\mathbf{k}}]$,
whose diagonal element gives the dispersion relations of the bulk bands.
The matrix $\mathcal{T}$ satisfies the para-unitary
\begin{eqnarray}\label{para2}
 \mathcal{T}^{\dag}\hat{\tau}\mathcal{T} &=&\hat{\tau},
\end{eqnarray}
where a diagonal matrix $\hat{\tau}$ takes $\pm1$ in the particle$/$hole space, i.e., $[\tau]_{jm} = \delta_{jm}\tau_j$ with $\tau_j = +1$ for $j =1,\dots,N$ and $\tau_j =-1$ for $N+1,\dots,2N$.  $N$ is the number of bands. The paraunitary defines the Berry Curvature of $\alpha$ bulk energy band given by,
\begin{eqnarray}
\nonumber \Omega_{ij;\alpha} &=&-2\textbf{Im}\left[ \hat{\tau}(\partial_{\emph{i}}\mathcal{T}^{\dg})\hat{\tau}(\partial_{\emph{j}}\mathcal{T})\right]_{\alpha\alpha}\\
\nonumber&=&-2\textbf{Im}\left[\tau_{\alpha\alpha}(\partial_{\emph{i}}\mathcal{T}^{\dg}_{\alpha})\hat{\tau}(\partial_{\emph{j}}\mathcal{T}_{\alpha})\right]\\
&=&-2\textbf{Im}\left[\tau_{\alpha\alpha}\langle\partial_{\emph{i}}\mathcal{T}^{\dg}_{\alpha}|\hat{\tau}\partial_{\emph{j}}\mathcal{T}_{\alpha}\rangle\right].
\label{Omega}
\end{eqnarray}
Using Eq.~\eqref{para1} and Eq.~\eqref{para2}, the completeness operator is derived as
\begin{eqnarray}\label{com}
\sum^{2N}_{\beta}\tau_{\beta\beta}|\hat{\tau}\mathcal{T}_{\beta}\rangle\langle\mathcal{T}^{\dag}_{\beta}|&=&\mathbf{I}_{2N\times 2N},\\
\sum^{2N}_{\beta}\tau_{\beta\beta}|\hat{\tau}\mathcal{T}_{\beta}\rangle\langle\mathcal{T}^{\dag}_{\beta}||\hat{\tau}\mathcal{T}_{\alpha}\rangle &=&|\hat{\tau}\mathcal{T}_{\alpha}\rangle.
\end{eqnarray}
Insert Eq.~\eqref{com} into Eq.~\eqref{Omega}, we have
\begin{align}\label{pDp}
\nonumber\langle\mathcal{T}^{\dg}_{\beta}|H|\mathcal{T}_{\alpha}\rangle &= \langle\mathcal{T}^{\dg}_{\beta}|E_{\alpha}\hat{\tau}|\mathcal{T}_{\alpha}\rangle, \\
\nonumber\langle\partial_{\emph{j}}\mathcal{T}^{\dg}_{\beta}|H|\mathcal{T}_{\alpha}\rangle+\langle\mathcal{T}^{\dg}_{\beta}|\partial_{\emph{j}}H|\mathcal{T}_{\alpha}\rangle
+\langle\mathcal{T}^{\dg}_{\beta}|H|\partial_{\emph{j}}\mathcal{T}_{\alpha}\rangle &=\langle\partial_{\emph{j}}\mathcal{T}^{\dg}_{\beta}|E_{\alpha}\hat{\tau}|\mathcal{T}_{\alpha}\rangle+\langle\mathcal{T}^{\dg}_{\beta}|\partial_{\emph{j}}E_{\alpha}\hat{\tau}|\mathcal{T}_{\alpha}\rangle+\langle\mathcal{T}^{\dg}_{\beta}|E_{\alpha}\hat{\tau}|\partial_{\emph{j}}\mathcal{T}_{\alpha}\rangle, \\
\nonumber\langle\mathcal{T}^{\dg}_{\beta}|\partial_{\emph{j}}H|\mathcal{T}_{\alpha}\rangle&=\langle\mathcal{T}^{\dg}_{\beta}|\hat{\tau}\partial_{\emph{j}}\mathcal{T}_{\alpha}\rangle(E_{\alpha}-E_{\beta}), \\
\langle\mathcal{T}^{\dg}_{\beta}|\hat{\tau}\partial_{\emph{j}}\mathcal{T}_{\alpha}\rangle &= \langle\mathcal{T}^{\dg}_{\beta}|\partial_{\emph{j}}H|\mathcal{T}_{\alpha}\rangle/(E_{\alpha}-E_{\beta}).
\end{align}
Then the Berry curvature can be written alternatively as:
\begin{align}\label{OmegaH}
\Omega_{ij;\alpha} =-2\textbf{Im}\left[ \tau_{\alpha\alpha}\sum^{2N}_{\beta\neq\alpha}\tau_{\beta\beta}\frac{\langle\mathcal{T}^{\dg}_{\alpha}|\partial_{\emph{i}}H|\mathcal{T}_{\beta}\rangle\langle\mathcal{T}^{\dg}_{\beta}|\partial_{\emph{j}}H|\mathcal{T}_{\alpha}\rangle}{(E_{\alpha}-E_{\beta})^2} \right].
\end{align}
We can immediately see that Eq.~\eqref{OmegaH} is manifestly gauge independent. It is more  advantageous to use Eq.~\eqref{OmegaH} to replace the Eq.~\eqref{Omega} as the Berry curvature does not depend explicitly on the phases of eigenvector. Obviously, the Berry curvature and the Chern number calculated by the Eq.~\eqref{OmegaH} are in accord with the relations,
\begin{align} \label{sum-rule}
\sum^{N}_{\alpha=1} C_\alpha =\sum^{2N}_{\alpha=N+1} C_\alpha = 0,\\
\Omega_{ij;\alpha}(\mathbf{k}) =-\Omega_{ij;\alpha+N}(-\mathbf{k})
\end{align}
as derived in Ref.~\cite{PhysRevB.87.174427}.
\twocolumngrid

\end{document}